\documentclass[12pt,preprint]{aastex}

\begin{document}
\title{A DENSE-CLOUD MODEL FOR GAMMA-RAY BURSTS\\
TO EXPLAIN BIMODALITY }
\author{F. Shekh-Momeni \footnote{e-mail: fmomeni@mehr.sharif.edu}
and J. Samimi \footnote{e-mail: samimi@sharif.edu}}

\affil{Department of Physics, Sharif University of Technology,\\
Tehran, P.O.Box: 11365-9161, Iran}

\begin{abstract}
In this model a collimated ultra-relativistic ejecta  collides
with an amorphous dense cloud surrounding the central engine,
producing gamma-rays via synchrotron process. The ejecta is taken
as a standard candle, while assuming a gaussian distribution in
thickness and density of the surrounding cloud. Due to the cloud
high density, the synchrotron emission would be an instantaneous
phenomenon (fast cooling synchrotron radiation), so a GRB duration
corresponds to the time that the ejecta takes to pass through the
cloud. Fitting the model with the observed bimodal distribution
of GRBs' durations, the ejecta's initial Lorentz factor, and its
initial opening angle are obtained as $\Gamma_{0}\lesssim
10^{3}$, and $\zeta_{0} \approx 10^{-2}$, and the mean density
and mean thickness of the surrounding cloud as $\overline{n} \sim
3 \times 10^{17} cm^{-3}$ and $\overline{L} \sim 2 \times 10^{13}
cm$. The clouds maybe interpreted as the extremely amorphous
envelops of Thorne-Zytkow objects. In this model the two classes
of long and short duration GRBs are explained in a unique frame.

\end{abstract} 

\section{INTRODUCTION}

Undoubtedly, gamma-ray bursts (GRBs) have remained to be one of
the most exiting, intriguing, and enigmatic astrophysical
phenomena since their mysterious discovery in the past several
decades (for a recent expository review of GRBs the interested
reader is referred to the excellent work by J. I. Katz
\cite{katz}). Although no two GRBs resemble each other and each
one has its own peculiarities which makes the problem of modeling
GRBs very difficult, the whole of GRBs reveals several interesting
features. Since the publication of the BATSE data \citep{1st}
which included the observation of over $200$ GRBs and revealed an
almost uniform distribution of the location of GRBs in the sky,
combined with the deficiency of faint GRBs, the association of
GRBs with the galactic plane has been ruled out. The successive
publications confirmed the figure more and more \citep{3th,4th}.
However, since the observation of afterglows in X-ray
\citep{Costa}, optical \citep{van Paradijs} and radio spectrum
\citep{Frail} and the advert of Robotic Optical Transient Search
Experiment (ROTSIE) telescope \citep[e.g.]{Akerlof,Gisler} which
has revealed the red shifts for several
GRBs, their cosmological origin is widely accepted.\\
The BATSE data \citep{1st,3th,4th} has also revealed another
equally important overall feature of the GRBs. The distribution
of time duration in observed GRBs shows a double heap
distribution, which the smaller one peaks around 0.2 sec and the
larger one peaks around 20 sec (Fig.1). This two peaked
distribution which apparently separated GRBs into the so called
short duration and long duration ones was referred to as
$\textit{bimodality}$ \citep{kouveliotou,Norris} and led some
investigators to believe that there are two distinct populations
of GRBs.\\
It has been widely believed that whatever the central engine is,
the radiation reaching us originates from the space surrounding
the central engine. It is also believed that during the collapse
the energy release streams out in relativistic confined ejectas
(not isotropically) and thus the total energy release during each
event is far less than the unbelievable amount that one might
obtain
by assuming isotropic radiation \citep{kulkarni}.\\
The aim of this paper is to present a rather simple model, based
on these general ideas, and show that there is no need for
assuming two distinct populations of GRBs, and to show that once
the geometrical considerations and cosmological effects are fully
accounted for a genetic standard candle ejecta, crossing an
amorphous dense cloud, the so called
$\textit{bimodality}$ can be deduced.\\
In \S~2 the model and its general formulation is introduced. In
\S~3 the computational results and the fit of model parameters
with BATSE data will be presented as well. \S~4 is devoted to a
discussion on the results. Some needed calculations and
discussions are presented in appendices.

\section{THE MODEL FORMULATION}
GRBs are modeled as a central engine with an instantaneous
ultra-relativistic jet of material surrounded by an amorphous
dense cloud. The central engine and its jet are taken as a
standard candle with a total release energy of $E$, and an initial
Lorentz factor $\Gamma_{0}$ and initial opening angle $\zeta_{0}$
of the jet for all GRBs, whereas the clouds are considered to have
a distribution both in thickness as well as in density. For the
sake of illustration and brevity we take both of these
distributions to be Gaussian.\\
We want to calculate the distribution of logarithm of time
duration of observed GRBs according to the above model. We should
however explore the ejecta evolution since the observed gamma ray
emission originates in the shock front of ejecta+shocked medium.
\subsection{The Ejecta Evoution}

The equations describing the ejecta evolution are presented here,
based on the notations of Paczynski \& Rhoads \cite{Paczynski and
Rhoads}. We consider the cloud to be at a distance $r_{0}$ from
the source, which may be negligible in comparison to the cloud
thickness $L$. The ratio of swept-up mass to ejecta mass $M_{0}$
has the form below:
\begin{equation}\label{}
f=\frac{1}{M_{0}}\int^{r}_{r_{0}} \rho \Omega_{m}(r^{\prime}) r
^{\prime 2} dr^{\prime},
\end{equation}
where, $r$ is the ejecta distance from the source. The cloud
density $\rho$ is taken to be independent of $r$. Furthermore,
$\Omega_{m}(r)=2\pi[1-\cos\zeta(r)]$, in which $\zeta(r)$
represents the opening angle of the ejecta at radius $r$. So,
equation(1) can be written as:
\begin{equation}\label{}
\frac{df}{dr}=2\pi n \frac{m_{p}\Gamma_{0}c^{2}}{
E}[1-\cos\zeta(r)]r^{2},
\end{equation}
where $E$ and $\Gamma_{0}$ are the initial kinetic Energy and
initial Lorentz factor of the ejecta ($E=\Gamma_{0} M_{0}
c^{2}$), and $n$ denotes the number density of the cloud
($\rho=m_{p}\:n$). Paczynski \& Rhoads \cite{Paczynski and
Rhoads} derived the relation between $f$ and $\Gamma$ from
conservation of energy and momentum. Here, we use their relation
in a form suitable for our computations:
\begin{equation}\label{}
\frac{df}{d\Gamma}=-\frac{\sqrt{\Gamma_{0}^{2}-1}}{\sqrt[3]{\Gamma^{2}-1}}~.
\end{equation}
The ejecta's opening angle $\zeta(r)$ increases with increasing
$r$ as a result of lateral spreading of the cloud of
ejecta+swept-up matter in the comoving frame at the sound speed
$c_{s}$, which has been derived by Rhoads \cite{Rhoads} to be as
below:
\begin{equation}\label{}
d\zeta(r)=\frac{c_{s}\:dt_{co}}{r},
\end{equation}
where $t_{co}$ denotes the time from the event, measured in the
ejecta comoving frame. Substituting $dt_{co} = dt/\Gamma$, and
$dt = dr / \beta c$, we have :
\begin{equation}\label{}
\frac{d\zeta(r)}{dr}=\frac{c_{s}/c}{\beta\:\Gamma \: r }.
\end{equation}
Now, eliminating $f$ between equations(2) and (3) yields:
\begin{equation}\label{}
\frac{d\Gamma}{dr}=-2\pi n
\frac{m_{p}\Gamma_{0}c^{2}}{E}[1-\cos\zeta(r)]
\frac{\sqrt[3]{\Gamma^{2}-1}}{\sqrt{\Gamma_{0}^{2}-1}}r^{2}~.
\end{equation}
Let's rewrite equations(5) and (6) in a non-dimensional form as
below:
\begin{equation}\label{}
\left\{\begin{array}{rcl}
\frac{d\zeta}{d\eta}&=&\frac{c_{s}/c}{\eta\Gamma(\eta)(1-\frac{1}{\Gamma^{2}})^{-1/2}}\\
\frac{d\Gamma}{d\eta}&=&-\frac{\Gamma_{0}\sqrt[3]{\Gamma^{2}-1}}{\sqrt{\Gamma_{0}^{2}-1}}
[1-\cos\zeta(\eta)]\eta^{2}
\end{array}\right. \nonumber\\,
\end{equation}
where, the non-dimensional parameter $\eta$ is defined as:
\begin{equation}\label{}
\eta\equiv \frac{r}{l(n/E)},
\end{equation}
in which:
\begin{equation}\label{}
l(n/E)\equiv\left(\frac{2\pi m_{p}c^{2}n}{E}\right)^{-1/3}=4.75
\times 10^{11}\left(\frac{n_{18}}{E_{48}}\right)^{-1/3}\:\:cm ~.
\end{equation}
These coupled first order differential equations can be solved
numerically by introducing the initial conditions:
\begin{equation}\label{}
\left\{\begin{array}{rcl}
   \Gamma(\eta_{0})=\Gamma_{0}\\
    \zeta(\eta_{0})=\zeta_{0}
    \end{array}\right. \nonumber\\,
\end{equation}
where $\eta_{0}\equiv r_{0}/l(n/E)$.\\
Noting that $\beta=dr/cdt=(1-1/\Gamma^{2})^{1/2}$, we have:
\begin{equation}\label{}
\frac{d\tau}{d\eta}=\left(1-\frac{1}{\Gamma^{2}(\eta)}\right)^{-1/2}~
~ ~ ; ~ ~ ~ \tau(\eta_{0})=0.
\end{equation}
In equation(11) we used equation(8) and a non-dimensional time
parameter $\tau$ defined as:
\begin{equation}\label{}
\tau\equiv\frac{c\:t}{l(n/E)}
\end{equation}
The numerical results of equation(11) are used in appendix C,
where we consider the effect of burster geometry on the
observable time duration.
\subsection{Formulation of Time Duration Distribution}
We begin with introducing the probability density for a collimated
burst to occur in a direction through the cloud with a thickness
$L$ and a number density $n$. We assume the cloud thickness to
have a gaussian distribution in various directions from the
central engine. By assuming a gaussian distribution for the cloud
density as well we have:
\begin{equation}\label{}
\frac{d^{3}p}{dn\:dL\:d\Omega}=\frac{1}{4\pi}\frac{1}{\sqrt{2\pi}\sigma_{_{n}}}
\:\exp\left[\frac{-(n-\overline{n})^{2}}{2\sigma_{_{n}}^{2}}\right]\:\frac{1}{\sqrt{2\pi}\sigma_{_{L}}}
\:\exp\left[\frac{-(L-\overline{L})^{2}}{2\sigma_{_{L}}^{2}}\right]~.
\end{equation}
The quantities $\overline{L}$ and $\sigma_{_{L}}$ denote the mean
thickness of the cloud and its dispersion, respectively. Likewise,
$\overline{n}$ and $\sigma_{_{n}}$ are defined in a similar
manner. Denoting the angle between the ejecta symmetry axis and
the line of sight by $\theta$ (Fig.2), and considering the
independence of the probability density from azimuth angle in
equation(13), we can write:
\begin{equation}\label{}
\frac{d^{3}p}{dn\:dL\:d\theta}=\frac{2\pi\:\sin\theta}{4\pi}\frac{1}{\sqrt{2\pi}\sigma_{_{n}}}
\:\exp\left[\frac{-(n-\overline{n})^{2}}{2\sigma_{_{n}}^{2}}\right]\:\frac{1}{\sqrt{2\pi}\sigma_{_{L}}}
\:\exp\left[\frac{-(L-\overline{L})^{2}}{2\sigma_{_{L}}^{2}}\right]~.
\end{equation}
The synchrotron emission is a fast process for our model (see
appendix A), so $T_{rec}$ (the time duration of a GRB as measured
by an observer cosmologicaly near to the source and located on
the line of sight), is attributed to the time that the shock front
takes to cross the dense cloud. As explained in appendix C, the
cloud thickness $L$ can be expressed as a function of $\theta$,
$n$, and $T_{rec}$ (Eqn.[C16]):
\begin{equation}\label{}
L=L(T_{rec},n,\theta)~.
\end{equation}
So we write:
\begin{equation}\label{}
\frac{d^{2}}{dn\:d\theta}\:\left(\frac{dp}{dL}\right)=\frac{d^{2}}{dn\:d\theta}
\left(\frac{dp}{dT_{rec}}\right)_{_{n,\theta}}\:\left(\frac{dT_{rec}}{dL}\right)_{_{n,\theta}}~,
\end{equation}
or:
\begin{equation}\label{}
\frac{d^{3}p}{dn\:d\theta\:d\log
T_{rec}}=\frac{d^{3}p}{dn\:d\theta\:dL}\left(\frac{dL}{d\log
T_{rec}}\right)_{_{n,\theta}}~,
\end{equation}
notifying that by this substitution(Eqn.[15]), the bursts that do
not manage to cross through the cloud with a thickness $L$ (and
stop in it) are practically omitted (see appendix C). Using
equation(14) in equation(17), it is seen that:
\begin{eqnarray}\label{}
\frac{d^{3}p}{dn\:d\theta\:d\log
T_{rec}}&=&\frac{\sin\theta}{4\pi}\frac{1}{\sigma_{_{n}}\:\sigma_{_{L}}}\left(\frac{dL}{d\log
T_{rec}}\right)_{_{n,\theta}}\nonumber\\
&\times&\:\exp\left[\frac{-(n-\overline{n})^{2}}{2\sigma_{_{n}}^{2}}\right]
\exp\left[\frac{-(L(T_{rec},n,\theta)-\overline{L})^{2}}{2\sigma_{_{L}}^{2}}\right].
\end{eqnarray}
Integrating over $\theta$ and $n$ yields:
\begin{eqnarray}\label{}
\frac{dp}{d\log T_{rec}}=\frac{1}{4 \pi\sigma_{_{n}}\sigma_{_{L}}}
\int_{\theta=0}^{\frac{\pi}{2}}\int_{n=0}^{\infty}
&&\hspace{-.7cm}\exp\left[\frac{-(n-\overline{n})^{2}}{2\sigma_{_{n}}^{2}}\right]
\exp\left[\frac{-(L(T_{rec},n,\theta)-\overline{L})^{2}}{2\sigma_{_{L}}^{2}}\right]\nonumber\\
&\times&\left(\frac{dL}{d\log
T_{rec}}\right)_{_{n,\theta}}\sin\theta\:d\theta\:dn~.
\end{eqnarray}
The effect of red shift is not considered yet. Equation(19) only
gives the probability density for an observed burst to have a
specified logarithm of time duration, as measured by an observer
near to it. We now investigate the relation between $dp / d\log
T_{rec}$ and $dp / d\log T_{\oplus}$ where $T_{\oplus}$ stands
for the time duration of a GRB measured at Earth. To obtain the
later, the former must be integrated over red shift z, using a
weight function $F_{_{GRB}}(z)$, so that $F_{_{GRB}}(z)\:dz$
represents the probability for occurring a GRB in a red shift
between $z$ and $z+dz$. To show this, let's consider an observer
located on the line from us to an occurred GRB which is
(cosmologicaly) near to it. The probability density for the GRB
to occur in a red shift z (with respect to us), and to have a
specific $\log T_{rec}$ (measured by the observer near to the
GRB), is clearly as below:
\begin{equation}\label{}
\frac{d^{2}p}{dz\:d\log T_{rec}}=F_{_{GRB}}(z)\frac{dp}{d\log
T_{rec}}~.
\end{equation}
To obtain $d^{2}p / dz\:d\log T_{\oplus}$, which is the
probability density for observing a GRB occurred at a red shift z
and observed to have a specific $T_{\oplus}$, we write:
\begin{equation}\label{}
\frac{d^{2}p}{dz\:d\log T_{\oplus}}=\frac{d}{d z}\left(\frac{d
p}{d \log T_{\oplus}}\right)_{z}=\frac{d}{d z}\left[\left(\frac{d
p}{d \log T_{rec}}\right)_{z}\left(\frac{d \log T_{rec}}{d \log
T_{\oplus}}\right)_{z}\right]~,
\end{equation}
noting that $T_{\oplus}=(1+z)T_{rec}$, the second term in the the
bracket equals one. Now, using equation(20) we have :
\begin{equation}\label{}
\frac{d^{2}p}{dz\:d\log
T_{\oplus}}=F_{_{GRB}}(z)\:\left(\left.\frac{dp}{d\log
T_{rec}}\right|_{T_{rec}=T_{\oplus}/(1+z)}\right)~,
\end{equation}
After integrating equation(22) over z, the final form of the
observable probability density will be as below:
\begin{equation}\label{}
\frac{dp}{d\log
T_{\oplus}}=\int_{z=0}^{\infty}F_{_{GRB}}(z)\:\left(\left.\frac{dp}{d\log
T_{rec}}\right|_{T_{rec}=T_{\oplus}/(1+z)}\right)\:dz\:~.
\end{equation}
The explicit form of $F_{_{GRB}}(z)$ is needed. This form is
obtained in appendix D (Eqn.[D4] and Eqn.[D5]). The second term in
the integrand is given by equation(19), in which the implicit
form of $L(T_{rec},n,\theta)$ appears. Appendix C is devoted to
the procedure of obtaining this function. For evaluation of
$dp/d\log T_{\oplus}$ (Eqn.[23]), we need the values of eight
parameters. Four of them are the cloud parameters $\overline{L}$,
$\sigma_{L}$, $\overline{n}$, and $\sigma_{n}$, and the fifth one
is the index $q$ corresponding to the GRB occurrence rate (see
Eqn.[D5]). The later three ones are the ejecta parameters, $E$,
$\Gamma_{0}$, and $\zeta_{0}$, which are the initial kinetic
energy, the initial Lorentz factor, and the initial opening angle
of the ejecta, respectively.
\section{NUMERICAL COMPUTATIONS AND RESULTS}
The $\textit{Mathmatica 4}$ software is used in the numerical
computations. In the procedure we begin with solving the coupled
differential equations(7) and (11) which govern the ejecta
evolution, and in which $\Gamma_{0}$ and $\zeta_{0}$ are the only
free parameters. Furthermore, we take $\eta_{0}=0$. For fixed
values of these parameters, the functions $\Gamma(\eta)$,
$\tau(\eta)$ and $\zeta_{rad}(\eta)$ (see Eqn.[C3]) can be
uniquely obtained . Obviously $\Gamma$ decreases with increasing
$\eta$ (Fig.3), and reaches to $\Gamma=1$ when $\eta$ approaches a
certain value. This in fact takes an infinite time and results in
an infinite non-dimensional time duration $\tau_{rec}$ (see
Eqn.[C4] and Eqn.[C5]). So, to circumvent the difficulty, we
consider an effective lower limit $\Gamma_{min}$ for the Lorentz
factor of shocked matter, which yields an effective upper limit
for non-dimensional time $\tau$. We adopted $\Gamma_{min}=2$ as a
lower cut-off. Then, following the procedure explained in appendix
C, for a specific cloud thickness $L$, and correspondingly a
specific $\eta_{_{L}}$ (Eqn.[C13]), the non-dimensional time
duration $\tau_{rec}(\theta,\eta_{_{L}})$ of a GRB can be
calculated for every $\theta$ and $\eta_{_{L}}$ (Fig.4). Let's
denote the radius corresponding to $\Gamma_{min}$ by $\eta_{m}$
so that $\Gamma_{min}\equiv\Gamma(\eta_{m})$, and recall that in
the model, for $\eta_{m}<\eta_{_{L}}$, the emitted photons would
be completely scattered by the electrons of not-swept part of the
cloud (that they have to pass through, before entering free space;
see appendix C), and so, the whole phenomenon may be called a
"$\textit{failed GRB}\:$". But, if $\eta_{m}>\eta_{_{L}}$, the
shocked matter succeeds to go out of the cloud and, as explained
in \S~4.3, due to a suppression process that intensively decreases
the cross-section of Compton scattering,(the major part of)the
emitted photons finally succeed to get released from the shocked
medium and enter free space, provided that
$\zeta(\eta_{_{L}})>[\sqrt{5 / 3}\:\Gamma(\eta_{_{L}})]^{-1}$
(see appendix B). It is really for this reason that $\tau_{rec}$
happens to be a function of $\eta_{_{L}}$, and practically
independent of $\eta_{m}$. Then, solving
$\tau_{rec}(\theta,\eta_{_{L}})$ for $\eta_{_{L}}$ numerically,
we can obtain the function $\eta_{_{L}}(\tau_{rec},\theta)$ which
is the equivalent non-dimensional form of expression (15). We
rewrite equation(19) in a non-dimensional form suitable for
numerical computations:
\begin{eqnarray}\label{}
\frac{dp}{d \log \tau_{rec}}=\frac{1}{2 \varrho\: \varsigma }
\int_{\theta=0}^{\frac{\pi}{2}}\int_{\nu=0}^{\infty}
&&\hspace{-.7cm}\exp\left[\frac{-(\nu-1)^{2}}{2\varrho^{2}}\right]
\exp\left[\frac{-\left(\frac{l(\overline{n}\nu/E)}{\overline{L}}\eta_{_{L}}(\tau_{rec},\theta)-1\right)^{2}}{2\varsigma^{2}}\right]\nonumber\\
&\times&\frac{l(\overline{n}\nu/E)}{\overline{L}}\left(\frac{d\eta_{_{L}}}{d\log
\tau_{rec}}\right)_{_{\nu,\theta}}\sin\theta\:d\theta\:d\nu,
\end{eqnarray}
in which, equation(C13) and the definitions below:\\
$\nu\equiv n / \overline{n}$~,\\
$\varrho\equiv \sigma_{n} / \overline{n}$~,\\
and\\
$\varsigma\equiv \sigma_{L} / \overline{L}$\\
are used. So, after choosing the quantities $\overline{L}$,
$\sigma_{_{L}}$, $\overline{n}$, $\sigma_{n}$, and of course $E$,
we can evaluate the integral appearing in equation(24). Then,
after choosing a value for $q$, the integral of equation(23) can
be performed to obtain the observable quantity $dp/d\textit{log} T_{\oplus}$.\\
As seen in equation(24), the initial kinetic energy $E$ and the
cloud mean density $\overline{n}$ appear only in the form
$E/\overline{n}$, and therefore they can not be found separately
in a fitting process, and all that can be obtained is only their
ratio. But, the observed fluence of GRBs reveals that the released
energy in GRBs is of order of $E_{iso}\sim10^{52} ergs$ for a
isotropic burst, which reduces to $(\Omega / 4\pi)E_{iso}$ if the
bursts were confined to a cone with a solid angle $\Omega$. We
used this amount of isotropic energy to relate $E$ to $\zeta_{0}$
($E\equiv E_{iso}\zeta_{0}^{2}/2 $), and reduce the free
parameters of the model to seven as
$\Gamma_{0},\zeta_{0},\overline{L},\sigma_{_{L}},\overline{n}$,
$\sigma_{n}$ and $q$, which hereafter are called $\Pi$
parameters. These parameters must be chosen so that the results
make the best fitting to the observed distribution of GRBs. To
achieve the task, one must search in the seven dimensional $\Pi$
space, and find the point in which the statistical quantity
$\chi^{2}$ takes the smallest value $\chi^{2}_{min}$. We used the
BATSE 4th catalogue \citep{4th} of 1234 GRBs and adopted the bins
$\Delta \log T _{\oplus}=0.2$ in a range $-1.9\leq\log
T_{\oplus}<2.9$, and used the $\textit{gradient search}$
technique to move toward the best point in which $\chi^{2}$ gets
minimized. The obtained fitted values are:
\begin{eqnarray}\label{}
\Gamma_{0}&=&0.97\times 10^{3} \:\:\:\:\:\:\:\:\:\:\:\zeta_{0} = 0.01 \nonumber\\
\overline{L}&=&1.7 \times 10^{13} cm \:\:\:\:\:\:\:\: \overline{n} = 2.9 \times 10^{17} cm^{-3}\nonumber\\
\sigma_{_{L}}&=&0.21\overline{L}\:\:\:\:\:\:\:\:\:\:\:\:\:\:\:\:\:\:\:\: \sigma_{n} = 0.71 \overline{n}\nonumber\\
q&=&-0.70
\end{eqnarray}
with a corresponding value $\chi^{2}_{min}=1.4$ (per degree of
freedom). As can be seen in Fig.5, the deviation from a more
perfect coincidence to data seems for the structure in the
observed distribution located around $\log T_{\oplus}\sim 0.7$.
The structure has been noted before \citep{Yu}. We put aside the
the data of the noted structure, which are ones with durations
between $0.3<\log(T_{\oplus})<0.9$, and again repeated the
numerical searching process in the parametric space. The new
obtained values of the fitted parameters are:\\
\begin{eqnarray}\label{}
\Gamma_{0}&=&0.96 \times 10^{3} \:\:\:\:\:\:\:\:\:\:\:\zeta_{0} = 0.01 \nonumber\\
\overline{L}&=&1.8 \times 10^{13} cm \:\:\:\:\:\:\:\: \overline{n} = 2.9 \times 10^{17} cm^{-3}\nonumber\\
\sigma_{_{L}}&=&0.21\overline{L}\:\:\:\:\:\:\:\:\:\:\:\:\:\:\:\:\:\:\:\: \sigma_{n} = 0.71 \overline{n}\nonumber\\
q&=&-0.70
\end{eqnarray}
which slightly differ from what previously obtained (Eqn.[25]).
But this time, $\chi^{2}_{min}$ reduces to $1.1$ (Fig.6). So the
mentioned structure may be interpreted as a result of an
independent phenomenon or effect which was not considered in our
model.

\section{DISCUSSION}
\subsection{The GRB Source}
Though the original shapes of equations (13) and (14) are
naturally
normalized, the resulting final equations (19) and (23) are not, because:\\
1)the bursts that their symmetry axes make an angle
$\theta>\zeta_{rad}(\eta_{_{L}})$ can not be detected (see appendix C),\\
2)the bursts for which $\eta_{m}<\eta_{_{L}}$ were not
considered in the numerical computations (because they do not manage to cross out the cloud).\\
The total probability of observing the occurred bursts,
$\int_{-3}^{3} (dp / d\log T_{\oplus}) ~ d\log T_{\oplus}$, is
obtained to be $1.47 \times 10^{-6}$, using our best fit
parameters(Eqn.[26]). This small probability must be interpreted
to be due to the above two reasons. The first reason describes the
suppression of observed GRBs by the term $ (\Omega / 4\pi)\sim
 \zeta_{0}^{2} / 2 \sim 10^{-4}$, while the second is responsible for the
remaining factor of $10^{-2}$. So, only about one percent of the
bursts manage to produce a real GRB, and only about $10^{-4}$ of
these GRBs occur in our line of sight. Of course, the lateral spreading of the ejecta+swept
mass may modify these two factors, increasing the first and decreasing the second.\\
With the values for the fitted parameters in equation(25) or in
equation(26), the mean mass of the clouds
$\overline{M}\equiv\frac{4}{3}\pi \overline{n}
m_{p}\overline{L}^{3}$ is about $1.2 \times10^{34}\:gr \approx 6
M_{\odot}$, which is of the order of the envelop mass in massive
stars. Such amorphous clouds seem strange in stars, but in close
neutron star-supergiant binaries where the neutron star orbits
around the core and accrete the envelope, the spherical symmetry
is likely removed, as pointed out by Podsiadlowski et al.
\cite{podsiadlowski}. Terman, Taam, \& Hernquist \cite{terman}
show that the system would emerge to form a red supergiant with a
massive Thorne-Zytkow Object (TZO) \citep{thorne}. Podsiadlowski
et al. \cite{podsiadlowski} also estimated a TZO birth-rate of
$\geq 10^{-4} yr^{-1}$ in the galaxy. Considering equation(D4) and
the obtained total probability of observing a "real" GRB, which is
$1.47\times 10^{-6}$ in our model (see above), it is seen that the
GRB observation rate ($2$ events per day) implies a total
(real+failed) GRB rate of the order of $\sim 10^{-3}Mpc^{-3}
yr^{-1}$. This is not too far from what Podsiadlowski et al.
\cite{podsiadlowski} theoretically estimated for TZO birth-rate
($\geq 10^{-4} galaxy^{-1} yr^{-1}$). Qin et al. \cite{qin}
introduced AICNS (Accretion-Induced Collapse of Neutron Stars)
scenario as GRB engines. Katz \cite{katz3} introduces a dense
cloud model to explain the observed $Gev$ gamma-rays in a number
of GRBs \citep[e.g.]{dingus,jones}, and suggests
the amorphous envelops of TZOs for these clouds.\\
The initial opening angle of the ejecta in our model
($\zeta_{0}\simeq 10^{-2}rad\approx 0.6 ^{\circ}$) is obtained
during the fitting process (\S~3). Such a small opening angle for
an ejecta or a jet might be explained by attributing it to the
collimating process of an ultra-relativistic ejecta with
$\Gamma_{0}>1/\zeta_{0}$ in a sufficiently high magnetic field
\citep{begelman}. The initial Lorentz factor of the ejecta, as
obtained in our model ($\Gamma_{0}\simeq 10^{3}$, see Eqn.[25])
provides the necessary condition of $\Gamma_{0}>1/\zeta_{0}$.
Aside from these theoretical justifications for the idea of a
highly collimated ejecta at the source, there are some pieces of
evidences supporting this idea \citep{lamb,waxman,granot2}.
\subsection{The Bimodality}
At the mean time the observed bimodality must be interpreted as a
result of the second reason expressed in \S~4.1. Though in our
model one may expect only one heap in the $log T_{\oplus}$
distribution associated with the directions in the clouds having
both the most probable $L$ and $n$ (which are equal or near to
$\overline{L}$ and $\overline{n}$), but as the numerical
computation shows, such directions in the clouds are too thick and
too dense to be crossed out by the ejecta, and therefore a real
GRB would
not be produced. So, we are left with four other regions with high probabilities:\\
1)  $n\sim\overline{n}$ and $L<\overline{L}$ (short GRBs)\\
2)  $n\sim\overline{n}$ and $L>\overline{L}$ (no GRBs)\\
3)  $n>\overline{n}$ and $L\sim\overline{L}$ (no GRBs)\\
4)  $n<\overline{n}$ and $L\sim\overline{L}$ (long GRBs)\\
Now, we show that the directions trough the clouds associated with
region(1) produce the short duration GRBs, and ones associated to
regions (2) and (3) produce no GRB, while the others in the forth
region produce the long duration ones.\\
As to the equation(9) the quantity $l(n/E)$ (which is of the
order of the sedov length) is $\sim 7\times 10^{11}cm \ll
\overline{L}$ for $n\sim \overline{n}$ (see Eqn.[25]). As seen in
Fig.7, the time duration of GRBs associated to such directions is
of the order of the time duration of short GRBs (Case (1) above).
Fig.8 shows that in our model the calculated time duration of
GRBs for $n\sim 10^{-6}\overline{n}$ is of the order of the
duration of long GRBs. Using equation(9), we see that in this
case ( $n\sim 10^{-6}\overline{n}$ ), $l(n/E)$ is about $5\times
10^{13}cm \sim \overline{L}$. So in our model the long duration
GRBs are due to the passing of ejecta through the directions
where $n \ll \overline{n}$ and $L \sim\overline{L}$ (case (4)
above). Furthermore, Since $l(n/E) \ll \overline{L}$ when
$n\sim\overline{n}$, we see that in cases (2) and (3) above, the
ejecta would stop in the dense cloud and the produced photons can
not scape from the optically thick cloud. In Fig.9,
$T_{rec}(L;n,\theta)$ is plotted for a number of densities, when
$\eta_{_{L}}=L / l(n/E)$ has the highest permitted value
$\eta_{m}$.\\
These general features of our calculations result from the
general features of our model, and therefore we speculate that
any distributions for the clouds thickness and density which are
picked around a mean value could explain the general features of
the duration distribution. Our chose of gaussian distributions
for thickness and density was only for the few parameters needed
to describe them.\\
\subsection{The Opacity}
In "The Dense Cloud Model" at this stage we have simply omitted
the bursts which ejectas can not go out of the cloud and stop in
it (because the emitted photons would be scattered by the dense
cloud), and we have claimed that the produced photons by all
bursts that succeed to cross out the dense cloud
can finally enter the free space.\\
At the first glance the model might appear to have a serious
problem in opacity, namely, in this model the best fit mean
density and mean thickness of the clouds are found to be of the
orders of $n\sim 10^{17}cm^{-3}$ and $L\sim 10^{13}cm$. So, as to
the relation $\tau_{op}=\sigma_{_{T}}\:n\:L$,
($\sigma_{_{T}}=6.65\times10^{-25}cm^{2}$) one would expect an
optical depth of the order of $10^{6}$. But there are two factors
that remedy the situation:\\
(i) As the numerical calculation shows (see \S~4.2), the most
probable directions characterized by $n\sim\overline{n}$ and
$L\sim\overline{L}$ are dynamically too thick to be crossed out
by the ejecta and the radiation produced in this case would be
completely scattered by the dense cloud. On the other hand, as
discussed in \S~4.2, the long duration GRBs are due to the
crossing of ejecta through directions where $n\sim 10
^{-6}\overline{n}$ and $L\sim\overline{L}$. since the density is
reduced by the factor $10^{-6}$, the optical depth drops to the
order of $1$. As explained in \S~4.2, the short duration GRBs are
due to the directions in the cloud where $n\sim\overline{n}$ and
$L\sim 10^{-2}\overline{L}$. In this case the optical depth of
the cloud reduces to $\sim10^{4}$, which is yet too high. But,\\
(ii) most of photons emitted off the shock front have the chance
to be overtaken by the moving shock (appendix B). Moreover, it has
been shown \citep{Momeni} that in high temperatures
$kT\sim10^{6}\:m_{e}c^{2}$ the cross section of Compton
scattering for instance for $Mev$ photons effectively drops to
$\sim 10^{-6}\:\sigma_{_{T}}$, so such a high temperature plasma
is much more transparent than what may seem at first. The same is
true for the case of a power law distribution of electrons (see
Eqn.[A4]). Most of the electrons in such a distribution have
energies of the order of $\gamma_{e,min}m_{e}c^{2}\sim 2 \times
10^{5}m_{e}c^{2}$, in which we have used equation(A5), taking
$p=2.5$, $\xi_{e}=1/3$ and $\Gamma=10^{3}$. For this case the
Compton cross section suppresses by the factor of
$\sim10^{-3}-10^{-4}$ for $100ev-1Kev$ photons (the energy of the
emitted photons in the shocked medium comoving frame is less than
their observed energy by the factor of $\Gamma$), and therefore,
the optical depth drops to the order of $1$. So the photons
overtaken by the shocked medium may remain in it without being
scattered until the shocked medium crosses up the dense cloud.
Briefly, due to this second factor, the optical depth in short
duration case diminishes to $\tau_{op}\sim1$, and the
optical depth in long duration case diminishes to $\tau_{op}\ll1$.\\
The variability seen in the light curves of long duration GRBs
maybe attributed to the heterogeny of cloud's density in the
ejecta's trajectory. This maybe the case in long duration GRBs
for which $\tau_{op}\ll1$, but in short duration GRBs for which
$\tau_{op}\sim1$, we expect
that such a heterogeny would not be appeared.\\
The lateral expansion of the shocked medium can reduce its
density and therefore its optical depth. This effect of course
may cool the shocked medium and cause an increase in effective
Compton cross section in it. The consideration of dynamical
feedback makes our modeling much more complicated and it has been
neglected here. In "The Dense Cloud Model" at this stage we have
simply omitted the bursts which ejectas can not go out of the
cloud and stop in it, but have claimed that the produced photons
by all bursts that succeed to cross out the
dense cloud can finally enter free space.\\

A complete study needs to include flux computations. An exact
comparison with observed time duration data would be possible
only when the theoretical time duration appearing in this work
were exactly evaluated in the same manner as the observable
duration $T_{90}$ is defined (as the time interval over which $5$
percent to $95$ percent of the burst counts accumulate).
Moreover, the BATSE's triggering mechanism made it less sensitive
to short GRBs than to long ones and therefore short GRBs were
detected to smaller distances \citep{mao,cohen,katz2}, so a
smaller number of them have been observed. Lee \& Petrosian
\cite{petrosian} studied this effect and corrected the number of
short GRBs. In a more exact work this correction must be
considered too.

We acknowledge the anonymous referee for valuable comments. F. S.
acknowledges Mr. Mehdi Haghighi for his guides on computational
methods. This research has been partly supported by Grant No.
NRCI 1853 of National Research Council of Islamic Republic of
Iran.
\newpage
\appendix

\section{Scyncrotron Cooling Time in Dense Cloud Model}

In this part we use the review paper of Piran (1999) to estimate
the synchrotron cooling time in our model. The synchrotron
cooling time in the comoving frame is:
\begin{equation}\label{}
t_{syn,co}=\frac{3m_{e}c}{4\sigma_{T}U_{_{B}}\:\gamma_{e}}~ ,
\end{equation}
where $\sigma_{T}$ is the Thompson cross section and $\gamma_{e}$
is the Lorentz factor of the emitting electron, while $U_{_{B}}$
stands for the energy density of magnetic field which in GRB
literature is assumed to be proportional to $u$, the comoving
internal energy density of the shocked matter:
\begin{equation}\label{}
U_{_{B}}=\xi_{_{B}}u ~ ,
\end{equation}
so that $\xi_{_{B}}$ represents the share of magnetic field in
$u$, which is given by:
\begin{equation}\label{}
u=4\Gamma^{2}n m_{p} c^{2}~,
\end{equation}
in which, $n$ is the number density of surrounding medium (ISM, or
a dense cloud as assumed in our model), and $\Gamma$ denotes
the Lorentz factor of the shocked matter.\\
The electrons in shocked media are assumed to develop a power law
distribution of Lorentz factors:
\begin{equation}\label{}
N(\gamma_{e})\propto\gamma_{e}^{-p}\:\:\:\:\:\:\:\:\:\:\:\:\:for
:\:\:\:\:\:\:\:\:\:\:\:\gamma_{e}>\gamma_{e,min}~.
\end{equation}
The convergence of total energy of the electrons requires the
power index $p$ to be greater than $2$, while the assumed lower
limit $\gamma_{e,min}$ is to prevent the divergence of electron
number density and is obtained to be:
\begin{equation}\label{}
\gamma_{e,min}=\frac{m_{p}}{m_{e}}\:
\frac{p-2}{p-1}\:\xi_{e}\:\Gamma ~ ,
\end{equation}
where $\xi_{e}\equiv u_{e} / u$ represents the share of the
electrons in the internal energy of shocked matter.\\
Furthermore, the time interval between emission of two photons
from the same point in the comoving frame, $\delta t_{co}$, and
the time interval $\delta t_{\oplus}$, which represents the time
interval of their successive arrival to a cosmologically distant
observer (at the earth), are related as below:
\begin{equation}\label{}
\delta t_{\oplus}=(1+z)\frac{\delta t_{co}}{2\Gamma}~.
\end{equation}
Equations(A1-A6) are adopted from Piran (1999). Now, by
substituting equations(A2),(A3), \& (A5) in equation(A1), and
then using equation(A6), the synchrotron cooling time measured by
a terrestrial observer turns out to be:
\begin{eqnarray}\label{}
t_{syn,\oplus}<\frac{3}{32}(1+z)\xi_{e}^{-1}\xi_{B}^{-1}
\left(\frac{m_{e}}{m_{p}}\right)^{2}\left(\frac{p-1}{p-2}\right)(\sigma_{{^T}}n c)^{-1}\Gamma^{-4}sec\nonumber\\
\sim
10^{-12}(1+z)\left(\frac{\xi_{e}}{0.1}\right)^{-1}\left(\frac{\xi_{B}}{0.1}\right)^{-1}
\left(\frac{n}{10^{18}}\right)^{-1}\Gamma^{-4}sec ~,
\end{eqnarray}
which is clearly much less than all observed time durations of
GRBs. So, the synchrotron emission in a dense cloud model must be
considered as an instantaneous process, and therefore the shock
front must be regarded as the emitting surface. This allows us to
attribute time duration of a GRB merely to the time
that the shock front takes to cross the dense cloud.\\

\section{The relation between Lorentz factors of
the shocked matter and the emitting surface}

Here, we want to find the relation between $\Gamma$ and
$\Gamma^{\prime}$ which are respectively the Lorentz factors of
the shocked matter and of the shock front (which is the emitting
surface in our model; see appendix A). In Fig.10, $\beta$ and
$\beta^{\prime}$ correspond to shocked matter and shock front
speeds, both measured in the (central) source frame. In the
shocked matter frame (Fig.11) the dense cloud which is seen to
have a density $\Gamma\:n$, moves toward the shocked medium with
a speed $\beta$, while being compressed to a density equal to
$4\Gamma n$. Consequently, as illustrated in Fig.11, the shocked
medium expands towards right with a speed $\beta^{\prime}_{co}$
which is in fact the speed of the shock front in the shocked
matter frame (compare Fig.10 and Fig.11). Considering the
conservation of nucleon number, it is seen that:
\begin{equation}\label{}
\beta^{\prime}_{co}=\frac{\beta}{4} ~ .
\end{equation}
Considering the relativistic summation of velocities, we have:
\begin{equation}\label{}
\beta^{\prime}=\frac{\beta^{\prime}_{co}+\beta}{1+\beta
\beta^{\prime}_{co}} ~ .
\end{equation}
Noting the relations $\Gamma=(1-\beta^{2})^{-1/2}$ and
$\Gamma^{\prime}=(1-{\beta^{\prime}}^{2})^{-1/2}$, the
substitution of equation(B1) in equation(B2) finally yields:
\begin{equation}\label{}
\Gamma^{\prime}\simeq\sqrt{\frac{5}{3}}\:\:\Gamma ~ ,
\end{equation}
The distance from the origin to the shock front, denoted by
$r^{\prime}$, can be obtained by integrating $\beta^{\prime}$
over $t$:
\begin{equation}\label{}
r^{\prime}=\int_{0}^{t}\beta^{\prime}\:c dt\:+r_{0}~.
\end{equation}
The results are shown in Fig.12. As seen, the difference between
$r$ and $r^{\prime}$ never exceeds one percent. This result
justifies the use of symbol $r$ (instead of
$r'$ )for the location of the shock front through out our formulation.\\
Now, let's consider a photon that is radiated from the shock front
(the emitting surface) moving with the Lorentz factor $\Gamma'$
(Eqn.[B3]), in a direction which makes an angle $\epsilon$ with
the velocity vector of the shocked matter (Fig.10). if :
\begin{equation}\label{}
\epsilon > \Gamma'^{-1},
\end{equation}
the photon would be overtaken by the shock front. The majority of
the emitted photons fulfill this condition when $\Gamma'\sim
10^{3}$ and $\zeta_{0}\sim 10^{-2}$ (see Eqn.[25]).

\section{Geometrical Considerations}

Here the effect of burster geometry on the observed time duration
is investigated. At first, as shown in Fig.13, we consider a
radiating segment on the shock front. The symmetry axis is
denoted by $z^{\prime}$. Noting the definition of $\epsilon$ in
the figure, it is seen that in addition to a radial component
$dr/dt$, the velocity vector of the segment must have a lateral
component $v_{lat}$ which is equal to:
\begin{equation}\label{}
v_{lat}=\frac{\epsilon}{\Gamma(\eta) \zeta(\eta)}~c_{s}~,
\end{equation}
as measured in the source frame. In equation(C1) the
non-dimensional radius $\eta$, as defined in equation(8), is used
instead of $r$, . Equation(C1) is obtained simply by using
equation(5) and assuming that the lateral speed of the segment in
a frame moving (only radially and) instantaneously along with the
segment, is the fraction $\epsilon / \zeta(\eta)$ of sound speed
$c_{s}$ (which is the lateral speed of the emitting surface at
its edges), and noting that the lateral speed in the source frame
is less than its corresponding value in the (radially
instantaneous) comoving frame by  the factor $1/\Gamma$. It is
well known that the radiation emitted by the segment is almost
confined to an angle $1/\Gamma$. In Fig.13 the axis of the
radiation cone emitted by the segment is denoted by
$z^{\prime\prime}$, which is parallel to the velocity vector of
the segment and, as seen in the figure, makes an angle
$\epsilon+\delta$ with $z'$ axis, where
$\delta=\tan^{-1}[\epsilon c_{s} / (\Gamma \beta c \zeta)]$. So
the radiation angle $\zeta_{rad}(\eta,\epsilon)$ (as depicted in
the figure) can be written as below:
\begin{equation}\label{}
\zeta_{rad}(\eta,\epsilon)=\epsilon+\tan^{-1}\left(\frac{\epsilon\:
c_{s}/c}{\Gamma(\eta)\:\beta(\eta)\: \zeta(\eta)
}\right)+1/\Gamma(\eta)~,
\end{equation}
and the total radiation angle, defined as the radiation angle at
the edge of the emitting surface, can be written as:
\begin{eqnarray}\label{}
\zeta_{rad}(\eta)\equiv\zeta_{rad}(\eta,\epsilon=\zeta(\eta))=\zeta(\eta)+\tan^{-1}\left(\frac{
c_{s}/c}{\Gamma(\eta)\:\beta(\eta)\: }\right)+1/\Gamma(\eta)~.
\end{eqnarray}
This angle represents the cone of space illuminated by the
emitting surface. Using the results of equation(7), the total
radiation angle can be evaluated for every "$\eta$". Having
defined the total radiation angle, we explore the effect of
burster geometry on its observed time duration $T_{\oplus}$. As
shown in Fig.2, the problem is studied in a spherical coordinate
system in which the central engine is taken as the origin, and the
line-of-sight as the polar axis $z$. Consider a photon emitted off
a point on the emitting surface (the shock front) with radial
coordinate $r$ and polar coordinate $\Theta$; and at an instance
t, measured in the source frame (the point is not shown in the
figure). The relation between $t$ and the photon arrival time
$t_{rec}$ to a (cosmologically) near observer is obtained by
Granot, Piran, \& Sari \cite{granot}. Making suitable for our
model, it is adjusted to the form below:
\begin{equation}\label{}
t_{rec}=t-\frac{r\:\cos\Theta-r_{0}}{c}~.
\end{equation}
In this equation the instance $t=0$ is defined as the time that
the ejecta collides with the dense cloud at $r=r_{0}$; while
$t_{rec}=0$ is the time that the (cosmologically) near observer
receives the photon emitted at $t=0$ from the point with
coordinates $r=r_{0}$ and $\Theta=0$. Defining:
\begin{equation}\label{}
\tau_{rec}\equiv \frac{c\:t_{rec}}{l(n/E)}~,
\end{equation}
and noting equation(8) and equation(12), we rewrite equation(C4)
in the form below:
\begin{equation}\label{}
\tau_{rec}=\tau-\eta\:\cos\Theta+\eta_{0}~.
\end{equation}
Multiplying equation(C4) by the cosmological time dilation term
$(1+z)$, results in the arrival time as maybe observed at the
earth:
\begin{equation}\label{}
t_{\oplus}=(1+z)t_{rec}
=(1+z)\left(t-\frac{r\:\cos\Theta-r_{0}}{c}\right)~,
\end{equation}
or equivalently:
\begin{equation}\label{}
\tau_{\oplus}=(1+z)\tau_{rec}
=(1+z)(\tau-\eta\:\cos\Theta+\eta_{0})~,
\end{equation}
in which we used the non-dimensional time duration
$\tau_{\oplus}$ defined as below:
\begin{equation}\label{}
\tau_{\oplus}\equiv \frac{c\:t_{\oplus}}{l(n/E)}~.
\end{equation}
Now, as shown in Fig.2, we consider a situation where the
ejecta's symmetry axes makes an angle $\theta$ with the line of
sight. The necessary condition that at least some photons of the
emitting surface are detected by the near observer is:
\begin{equation}\label{}
\zeta_{rad}(\eta)>\theta ~.
\end{equation}
Here we denote the inverse of function $\zeta_{rad}(\eta)$ by
$\eta_{rad}(\zeta_{rad})$, which gives the radius corresponding to
$\zeta_{rad}$. As seen in Fig.2, for $\theta$'s larger than
$\zeta_{rad}(\eta_{0})$, the first photons reaching the detectors
are those emitted at $\eta=\eta_{rad}(\theta)$. So, we use
$\tau(\eta)$ (which is obtainable by solving equation(11)) to
write:
\begin{equation}\label{}
\tau_{1}(\theta)=\cases{0&\mbox if $\:\:\:\:\:\theta <
\zeta_{rad}(\eta_{0})$ \cr
              \tau(\eta_{rad}(\theta)) &\mbox if$\:\:\:\:\:\:\theta > \zeta_{rad}(\eta_{0})$},
\end{equation}
where $\tau_{1}(\theta)$ represents the starting time (in the
source frame) that the emitted photons can reach the
(cosmologicaly near) observer. Now, using the numerical results
of equation(11) we can obtain the function $\eta(\tau)$. Then,
considering equation(C6), the non-dimensional time
$\tau_{rec,1}(\theta)$ corresponding to $\tau_{1}(\theta)$ would
be as below:
\begin{equation}\label{}
\tau_{rec,1}(\theta)=\cases{0&\mbox :$\:\:\theta < \zeta_{0}$ \cr
               -\eta_{0}\cos(\theta-\zeta_{0})+\eta_{0} &\mbox :$\:\:\zeta_{0} < \theta < \zeta_{rad}(\eta_{0})$ \cr
                \tau(\eta_{rad}(\theta))-\eta_{rad}(\theta)\cos[\theta-\zeta(\eta_{rad}(\theta))]+\eta_{0}&\mbox :$\:\:\theta > \zeta_{rad}(\eta_{0})$
                }~.
\end{equation}
Now, we are to find the time $\tau_{rec,2}(\theta)$ after which
no photons can be detected by the near observer. In Fig.2, the
photons emitted from the edge point $A$ can reach us at all times
greater than $\tau_{1}(\theta)$. Let's remind that in our model
the emission process terminates at the time that the shocked
matter goes out of the cloud (appendix A). As is seen in
equation(C6), the closer to the point $B$ is a point on the
emitting surface, the later its emitted photons would reach the
near observer, of course, provided that the observer line-of-sight
remains in the radiation cone of the emitting point.\\
Defining:
\begin{equation}\label{}
\eta_{L}\equiv\frac{r_{0}+L}{l(n/E)}~,
\end{equation}
and recalling equation(C2), we can solve the equation:
\begin{equation}\label{}
\zeta_{rad}(\eta_{_{L}},\epsilon)=\theta~,
\end{equation}
to find the function $\epsilon=\epsilon(\eta_{_{L}},\theta)$,
which specifies the furthest point (on the shock front at the
radius $\eta_{_{L}}$) which its radiation reaches us, of course,
if it
remains smaller than $\zeta(\eta_{_{L}})$.\\
Now, using equation(C6), we can find the instance that the last
photons reach the near observer:
\begin{equation}\label{}
\tau_{rec,2}(\theta,\eta_{_{L}},\eta_{0})=\tau(\eta_{_{L}})-
\eta_{_{L}}\:\cos(\theta+min[\epsilon(\eta_{_{L}},\theta),\zeta(\eta_{_{L}})])+\eta_{0}~.
\end{equation}
Finally, the non-dimensional time duration of a GRB,
$\tau_{rec}(\theta,\eta_{_{L}})$, will be equal to
$\tau_{rec,2}(\theta,\eta_{_{L}})-\tau_{rec,1}(\theta)$ and, as
to equation(C12) and equation(C15), besides being a function of
parameters of the ejecta and the cloud, it is also a function of
the inclination angle $\theta$. So, the time duration of a GRB as
measured by an observer cosmologivally near to it would be a
function of $L$, $n$, and $\theta$:
\begin{equation}\label{}
T_{rec}=T_{rec}(L;n,\theta)~,
\end{equation}
where:
\begin{equation}\label{}
T_{rec}=\tau_{rec}(\theta,\eta_{_{L}})
\frac{l\left(n/E\right)}{c},
\end{equation}
The expression(C16) needs some explanations. In our model the
ejecta parameters $\zeta_{0}$ , $E$ and $\Gamma_{0}$ and the
radius $r_{0}$ are assumed to be the same in all GRBs. So, these
parameters do not appear in equation(C16) explicitly , though it
is implicitly a function of them too. At the mean time, the
cloud's density $n$ and its thickness $L$ are assumed to be
different in different directions, and therefore they do appear
explicitly in the expression (C16). Furthermore, if a cloud
thickness $L$ is much more than its associated $Sedov$ length
$l_{sedov}\equiv(E/n\:m_{p}\:c^{2})^{1/3}$, the ejecta may not
cross it up, and finally stops in it. In such a situation the
time duration of a GRB would not be a function of $L$, while it
still remains dependent on $n$ and $\theta$. This is why in
expression (C16), the quantity $L$ is distinguished from
$n$ and $\theta$ by a semicolon. The rearrangement of the expression(C16)
in the form $L=L(T_{rec},)\theta,n$ would be meaningful only when
we are dealt with the situation where the ejecta succeed to cross up the cloud (see Eqn.[15]).\\

\section{The explicit form of $F_{_{GRB}}(z)$}

Here the relation between $F_{_{GRB}}(z)$ appearing in
equation(23) and the GRB occurring rate $f_{_{GRB}}(z)$ (in units
of $Mpc^{-3}\:yr^{-1}$) is derived, so that by adopting a
cosmological model for the occurring rate, the integral in
equation(23) can be evaluated.\\
The number of GRBs that their effects could reach us in a time
interval $\delta t_{0}$ (which is very much less than the present
comoving time $t_{0}$), and from a spherical volume element
$\delta V_{z}$ (disregarding various effects, such as the
geometrical ones discussed in appendix C, or those related to
detectors threshold which affect the number of detected GRBs) is
as below:
\begin{eqnarray}\label{}
\delta^{2}N_{_{GRB}}=f_{_{GRB}}(z)\:\delta V_{z}\:\:\frac{\delta t_{0}}{1+z}\nonumber\\
=f_{_{GRB}}(z).\:4\pi R^{^{3}}(z)\:r^{^{2}}(z)\:\delta r(z)\:
\frac{\delta t_{0}}{1+z}~,
\end{eqnarray}
where $r(z)$ denotes the non-dimensional radial parameter of the
source that its effects reach us with a red shift z, and $R(z)$,
the scale factor at this red shift. In Einstein-de Sitter model
we have:
\begin{equation}\label{}
R_{0}\:r(z)=\frac{2c}{H_{0}}\{1-(1+z)^{-1/2}\}~,
\end{equation}
where $H_{0}$ is the Hubble constant. Furthermore, in FRW metrics:
\begin{equation}\label{}
R(z)=R_{0}(1+z)^{-1}~,
\end{equation}
where $R_{0}\equiv R(z=0)$. Now, using equations(D2) and (D3) in
equation(D1) we obtain:
\begin{equation}\label{}
F_{_{GRB}}(z) \propto \frac{\delta^{2}N_{_{GRB}}}{\delta t_{0}
\delta
z}=2\pi\:\left(\frac{2c}{H_{0}}\right)^{^{3}}(1+z)^{^{-11/2}}\:\{1-(1+z)^{-1/2}\}^{^{2}}\:f_{_{GRB}}(z)~.
\end{equation}
What is remained is the explicit form of $f_{_{GRB}}(z)$. The high
variability seen in GRB light curves has convinced the
investigators to relate GRBs to stellar objects , and
consequently their rate $f_{_{GRB}}(z)$ to the star formation
rate $f_{_{SF}}(z)$. The simplest model is of course a
proportional model $f_{_{GRB}}(z)\propto f_{_{SF}}(z)$. the
proportional model may be correct if GRBs are attributed to the
evolution of massive stars whose lifetime is negligible in
comparison with the cosmological time scale, but in NS-NS mergers
model the proportionality may not be valid (because of the delay
time from the star formation to NS-NS mergers). Wijers et al.
\cite{wijers} claimed that there is a good consistency between
the proportional model and the observed GRB brightness
distribution, while Petrosian \& Lloyd \cite{petrosian2} concluded
that none of the NS-NS and the proportional model are in
agreement with the observed $f_{_{SF}}(z)$. Totani \cite{totani}
ascribed this discrepancy to the uncertainties in SFR
observations. Anyway we simply assume the GRB rate to be as below:
\begin{equation}\label{}
f_{_{GRB}}(z)=f_{_{GRB}}(0)(1+z)^{^{3+q}}~,
\end{equation}
and treat $q$ as a free parameter that its best value should be
obtained during the fitting procedure. In Fig.14, $F_{_{GRB}}(z)$
is plotted for a number of $q$'s. Clearly the case $q=0$
corresponds to a universe where the changes in the rates of
astrophysical phenomena are only due to its expansion
(non-evolutionary universe). It can be seen that $F_{_{GRB}}(z)$
takes its maximum at $z\sim1$, which is not very sensitive to the
magnitude of $q$.

 \newpage
\begin{figure}
\centering
\includegraphics[width=8cm]{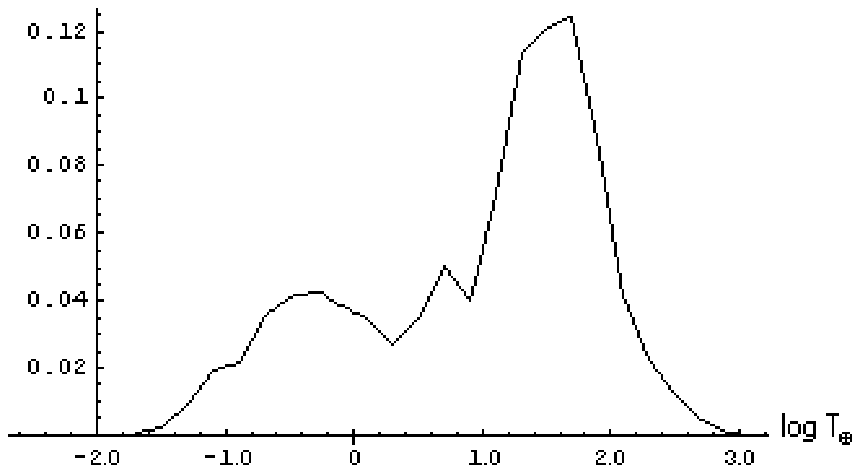}
      \caption{Distribution of log of GRBs' time Duration (normalized), obtained from data of BATSE 4th catalog.}
       \label{fit}
   \end{figure}
   \begin{figure}
   \centering
   \includegraphics[width=12cm]{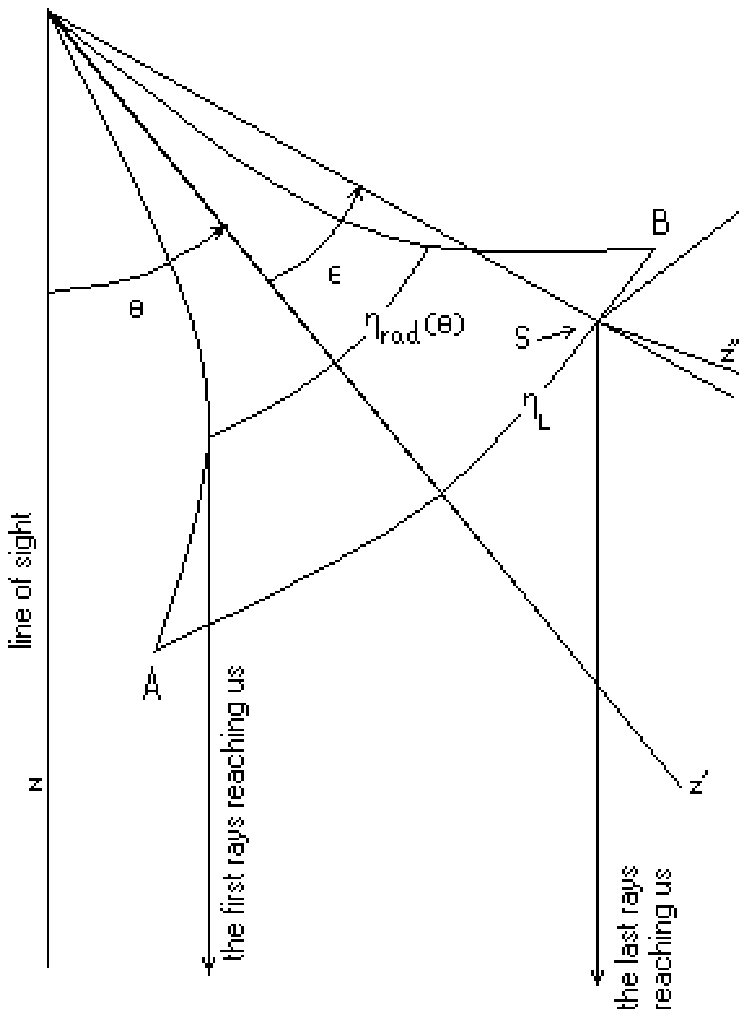}
      \caption{Geometry of Radiation. In the figure, $\eta_{rad}(\theta)$ denotes the radius where the
       radiation angle $\zeta_{rad}$ becomes equal to $\theta$ (see appendix C). As shown
       in the figure (for $\theta$'s larger than $\zeta_{rad}(\eta_{0})$) the first photons
       reaching us are those emitted from the edge of the shock front at this radius. If
       $\epsilon(\eta_{_{L}},\theta)<\zeta(\eta_{_{L}})$, then the last photons reaching
       us are those emitted from $\epsilon=\epsilon(\eta_{_{L}},\theta)$ on the shock front
       (point S in the figure). The point S would be the furthest visible point on the shock
        front, since the line of sight fells out of the radiation cone of the points farther
        than it (points between S and B). If $\epsilon(\eta_{_{L}},\theta)>\zeta(\eta_{_{L}})$,
        then the point B would be the last visible point (see Eqn.[C15]).}
       \label{fit}
   \end{figure}
   \begin{figure}
   \centering
   \includegraphics[width=8cm]{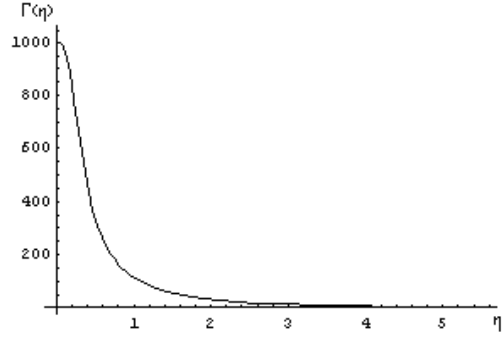}
      \caption{The Lorentz factor of shocked medium $\Gamma(\eta)$ versus the non-dimensional
      radius $\eta$, with $\Gamma_{0}= 1000$ and $\zeta_{0}=0.01$ (see Eqn.[7]).  }
       \label{fit}
   \end{figure}
   \begin{figure}
   \centering
   \includegraphics[width=12cm]{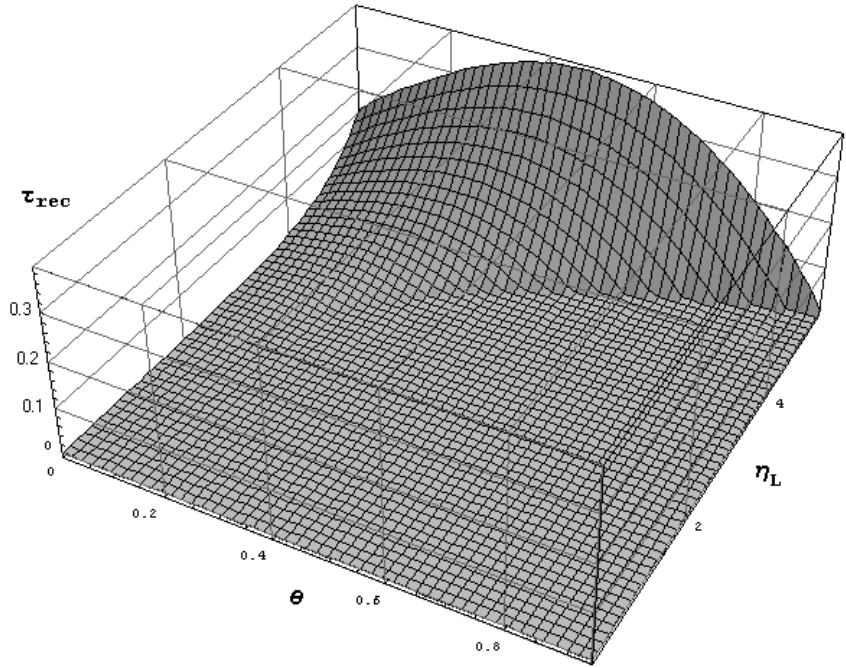}
      \caption{The evaluated non-dimensional time duration $\tau_{rec}(\theta,\eta_{_{L}})$
      of GRBs, versus $\theta$ and $\eta_{_{L}}$, with $\Gamma_{0}= 1000$ and $\zeta_{0}=0.01$,
      (see appendix C).}
       \label{fit}
   \end{figure}
   \begin{figure}
   \centering
   \includegraphics[width=8cm]{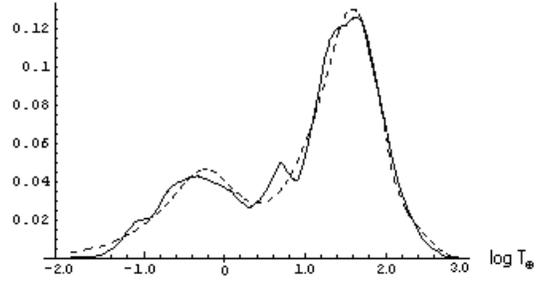}
      \caption{Results of best fitting. The solid curve shows the
       observed time duration distribution and the dashed one is
       calculated using the best of values for the model parameters. $\chi^{2}_{min}=1.4 pdf$ (see \S~3).}
       \label{fit}
   \end{figure}
\begin{figure}
   \centering
   \includegraphics[width=8cm]{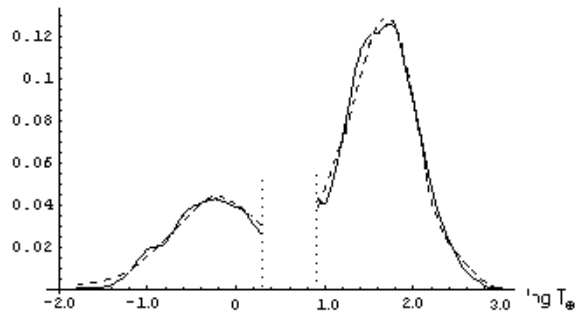}
      \caption{Results of best fitting. Same as Fig.5 with the data
      corresponding to the structure with $0.3 < \log T _{\oplus} < 0.9$ were cut. $\chi^{2}=1.1 pdf$ (see \S~3).}
       \label{fit}
   \end{figure}
\begin{figure}
   \centering
   \includegraphics[width=8cm]{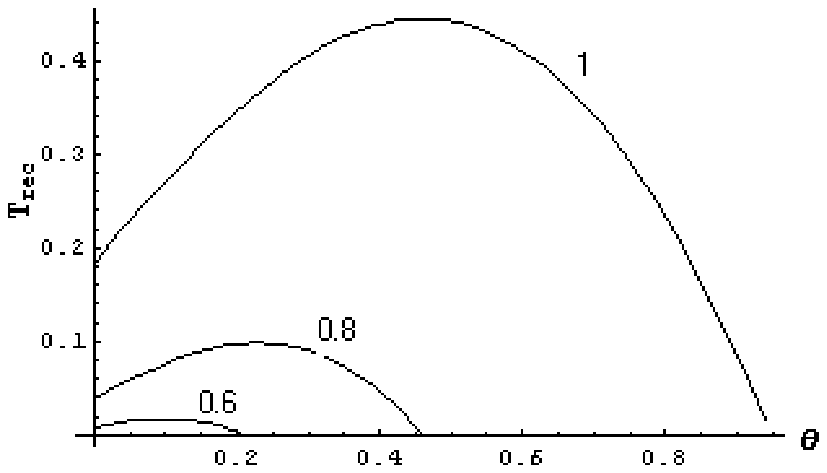}
      \caption{The time duration of GRBs, evaluated in our model
      with $n=\overline{n}$. It is plotted versus the inclination angle
$\theta$ for
      a number of possible thicknesses $\eta_{_{L}}$. As seen,
      the order of time durations are of the order of ones in short duration GRBs (see \S~4.2). }
       \label{fit}
   \end{figure}
\begin{figure}
   \centering
   \includegraphics[width=8cm]{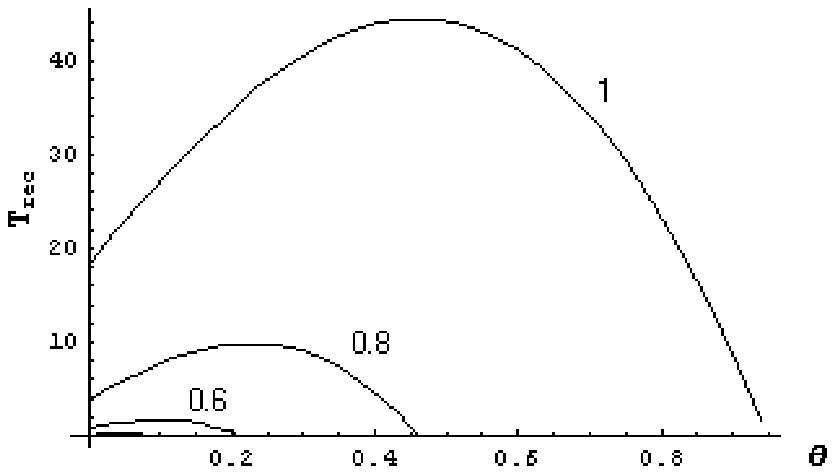}
      \caption{The time duration of GRBs, evaluated in our model
      with
      $n=10^{-6}\overline{n}$. It is plotted versus the inclination angle $\theta$
      for a number of possible thicknesses $\eta_{_{L}}$. As seen,
      the order of time durations are of the order of ones in long duration GRBs (see \S~4.2). }
       \label{fit}
   \end{figure}
\begin{figure}
   \centering
   \includegraphics[width=8cm]{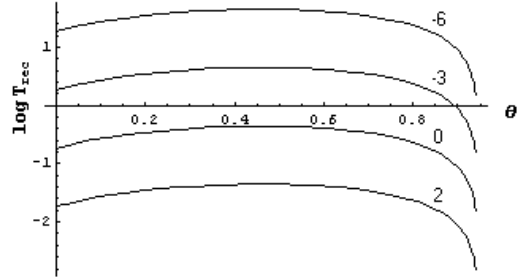}
      \caption{$\log T_{rec}(L;n,\theta)$ is plotted versus the inclination angle $\theta$
       for a number of densities, with $\eta_{_{L}}\equiv L / l(n/E)$ equal to the highest
       permitted value $\eta_{m}$. The number near each curve is $\log (n /
       \overline{n})$ ($\overline{n}=2.9 \times 10^{17} cm^{-3} $) (see \S~3 and \S~4.2 ). }
       \label{fit}
   \end{figure}
   \begin{figure}
   \centering
   \includegraphics[width=8cm]{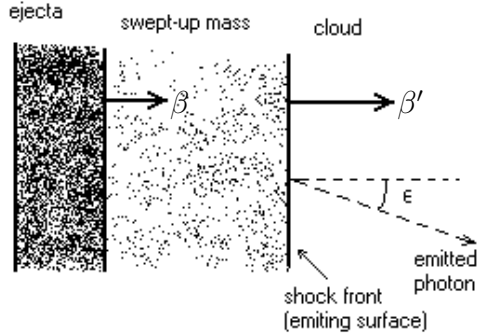}
      \caption{The shocked matter and the shock front (emitting surface) speeds,
      as viewed in the source frame. In the figure, if the angle $\epsilon$ were
larger than $1/\Gamma'$
      the photon would be overtaken by the shock front (from which it were emitted), so that the
       photon would remain in the shocked medium (see appendix B).}
\begin{picture}(100,100)(0,0)
\put(110,270){$\beta^{\prime}$} \put(24,270){$\beta$}
\end{picture}
\label{fit}
\end{figure}
\begin{figure}
   \centering
   \includegraphics[width=8cm]{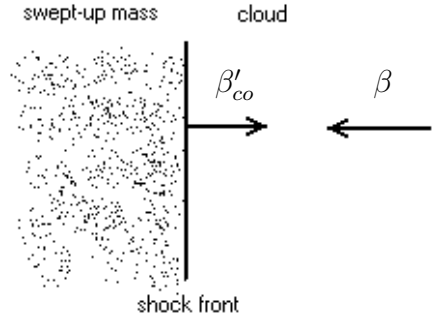}
      \caption{The shock front (emitting surface) speed $\beta'_{co}$, measured
      in the shocked frame. In this frame the cloud has a density of $\Gamma n$ and moves toward
      the shocked medium with a speed $\beta$, and becomes compressed to a density of $4 \Gamma n$.
       So clearly we have $ \beta'_{co} = \beta / 4$ (see appendix B).}
       \begin{picture}(100,100)(0,0)
\put(90,275){$\beta$} \put(30,275){$\beta^{\prime}_{co}$}
\end{picture}
       \label{fit}
   \end{figure}
   \begin{figure}
   \centering
   \includegraphics[width=8cm]{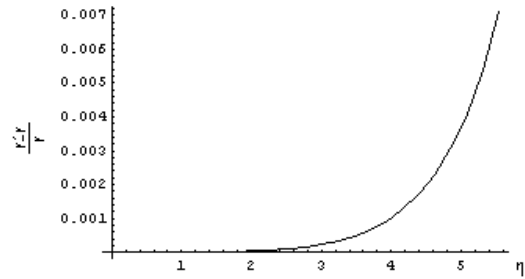}
      \caption{ $(r^{\prime} - r)/r$ versus $\eta \equiv r/l(n/E)$,
      with
      $ n = 2.9 \times 10^{17} cm^{-3}$ (see appendix B). }
       \label{fit}
   \end{figure}
   \begin{figure}
   \centering
   \includegraphics[width=12cm]{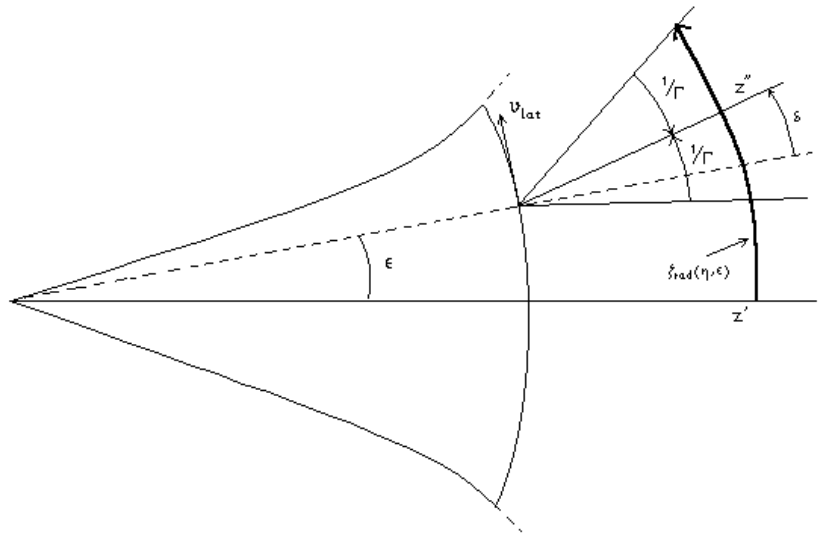}
      \caption{Geometry of Radiation. The lateral speed of the emitting segment, $v_{lat}$, causes the radiation cone
axis $z''$ to make an angle
      $\epsilon+\delta$ with $z'$ axis, where $\delta = \tan^{-1}[\epsilon
c_{s} / (\Gamma \beta c \zeta)]$ (see appendix C). }
       \label{fit}
   \end{figure}
\begin{figure}
   \centering
   \includegraphics[width=12cm]{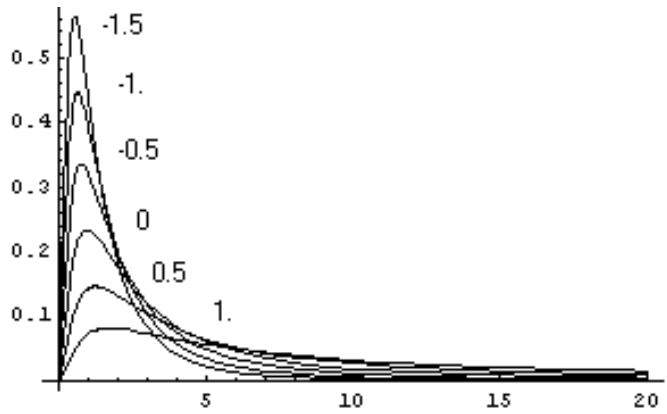}
      \caption{The quantity $F_{_{GRB}}(z)$ (the probability density of observing a GRB
      at a redshift $z$) is plotted for a number of $q$'s (the
      value of $q$
      is written near to the peak of its corresponding curve). As seen, all curves have
      a maximum at $z \sim 1$. This explains why the most of GRBs have redshifts $z
\sim 1$ (see Eqn.[D5]).}
       \label{fit}
   \end{figure}

\end{document}